\documentstyle[prb,epsfig,aps,preprint]{revtex}
\input{epsf}

\begin{document}
\draft
\title{Observation of progressive motion of ac-driven solitons}
\author{Alexey V. Ustinov$^{a}$ and Boris A. Malomed$^{b}$}
\address{$^{a}$Physikalisches Institut III, Universit\"at\\
Erlangen-N\"urnberg, D-91058, Erlangen, Germany\\
$^{b}$Department of Interdisciplinary Studies, Faculty of\\
Engineering, Tel Aviv University, Tel Aviv, Israel}
\date{\today}
\maketitle

\begin{abstract}
We report the first experimental observation of ac-driven phase-locked
motion of a topological soliton at a nonzero average velocity in a
periodically modulated lossy medium. The velocity is related by a resonant
condition to the driving frequency. The observation is made in terms of the
current-voltage, $I(V)$, characteristics for a fluxon trapped in an annular
Josephson junction placed into dc magnetic field. Large zero-crossing
constant-voltage steps, exactly corresponding to the resonantly locked
soliton motion at different orders of the resonance, are found on the
experimental $I(V)$ curves. A measured dependence of the size of the steps
vs. the external magnetic field is in good agreement with predictions of an
analytical model based on the balance equation for the fluxon's energy. The
effect has a potential application as a low-frequency voltage standard.
\end{abstract}


%\wideabs{

%}

%end of WideAbs
%\newpage

The important role played by nonlinear excitations in the form of solitons
in various physical systems is commonly known. However, experimental
observation of dynamical effects produced by solitons is often difficult
because real systems may be far from their idealized mathematical models
which give rise to soliton solutions. Among perturbations that destroy
soliton effects, dissipation is the most important one. To compensate
dissipative losses and thus make the soliton dynamics visible, one must
apply an external force that supports (in particular, {\em drives}) a
soliton \cite{oldreview}.

Solitons of the simplest type are topological {\it kinks}, a well-known
example being a magnetic flux quantum (fluxon) in long Josephson junctions
(LJJs) \cite{Barone,Ust-rev98}. A fluxon in LJJ can be easily driven by bias
current applied to the junction. The motion of a fluxon gives rise to dc
voltage $V$ across the junction, which is proportional to the fluxon's mean
velocity. Varying the dc bias current $I$, one can produce a dependence $%
V(I) $, which is the main dynamical characteristic of LJJ. An experimentally
obtained $I(V)$ curve easily allows one to identify the presence of one or
more fluxons trapped in LJJ \cite{Barone}.

Microwave field irradiating LJJ gives rise to an ac drive acting on the
fluxon. In a spatially homogeneous lossy system, an ac drive may only
support an oscillatory motion of a kink, which is hard to observe in LJJ due
to the absence of dc voltage. However, it was predicted \cite{Bob1} that an
ac drive can support motion of a kink with a {\em nonzero} average velocity $%
u$ in a system with a periodic spatial modulation. Indeed, a moving kink
passes the modulation length (period) $L$ during the time $L/u$. If this
time is commensurate with the period $2\pi /\omega $ of the ac drive, i.e., $%
m(L/u)=2\pi /\omega $ with an integer $m$, or
\begin{equation}
u=m(L\omega /2\pi ),  \label{spectrum}
\end{equation}
one may expect a resonance (of order $m$) between the two periodicities. In
other words, a moving kink may be phase-locked to the ac drive. This
provides for permanent transfer of energy from the drive to the kink, making
it possible to compensate dissipative losses. The energy balance gives rise
to a minimum ({\it threshold}) value $\Gamma _{{\rm thr}}$ of the ac drive's
amplitude $\Gamma $, which can compensate the losses and support the motion
of the ac-driven kink.

A more general case, when the system is simultaneously driven by the ac
field and dc bias current $I$, was considered in Ref.~\onlinecite{Bob2}. It was
predicted that the corresponding $V(I)$ characteristic has {\it steps}
(constant-voltage segments) at the resonant velocities (\ref{spectrum}). The
motion of the fluxon under the action of pure ac drive then corresponds to
{\it zero-crossings}, when the steps cross the axis $I=0$. In fact, the most
straightforward way to observe the ac-driven motion of the soliton is
through zero-crossings on the $V(I)$ characteristic.

A formally similar feature is known in{\em \ small} Josephson junctions as
Shapiro steps \cite{Barone}: ac drive applied to the junction gives rise to
dc voltage across it (an inverse ac Josephson effect). However, a drastic
difference of the effect sought for in this work from the Shapiro steps is
that they are only possible at high frequencies exceeding the junctions's
plasma frequency, while the ac-driven motion of the fluxon can be supported
by the ac drive with an arbitrarily low frequency. This circumstance also
opens way for application to the design of voltage standards using easily
accessible sources of low-frequency radiation, which are not usable with the
usual voltage standards based on small Josephson junctions. Moreover,
because the single-valued dc voltage in LJJ is controlled not only by
the frequency and amplitude of the ac drive, but also by locally applied magnetic
field (see below), the system studied here may be useful for designing an ac voltage
standard.

The inverse Josephson effect was also studied in detail in ac-driven
finite-length linear junctions with reflecting edges, corresponding to zero
external magnetic field \cite{Salerno}. A fluxon turns into an antifluxon
while bouncing from the edge. In that case, the system of constant-voltage
steps is complicated, on the contrary to a very simple set of two symmetric
steps in the circular junction, see below. This is explained by the fact
that the {\em shuttle motion} of the polarity-reversing fluxon/antifluxon in
the linear junction is different from the progressive motion of the
fluxon, without any polarity reversal, in the circular LJJ.

An objective of this paper is to report direct experimental observation of
the ac-driven fluxon motion in periodically modulated LJJs. Frequently, it
is assumed that the necessary periodic spatial modulation along the junction
can be induced by periodically changing the thickness of the dielectric
barrier separating two superconductors. In the presence of the losses and
drive, the modulated LJJ is described by the perturbed sine-Gordon (sG)
equation,
\begin{equation}
\phi _{tt}-\phi _{xx}+\left( 1+\varepsilon \sin \frac{2\pi x}{L}\right) \sin
\phi =-\alpha \phi _{t}-\gamma -\Gamma \sin \omega t,  \label{sG1}
\end{equation}
where $x$ and $t$ are the length along the junction and time, measured,
respectively, in units of the Josephson length and inverse plasma frequency,
$\varepsilon $ is the normalized modulation amplitude, while $\gamma $ and $%
\Gamma $ are the dc and ac bias current densities, both normalized to the
junction's critical current density.

The harmonic modulation of the local magnitude of the maximum Josephson
current, assumed by this model, is very hard to realize in an experiment due
to the exponential dependence of the critical current on the thickness of
the dielectric barrier. A much easier and fully controllable way to induce a
harmonic periodic modulation is to use an {\em annular} (ring- shaped) LJJ,
to which uniform dc magnetic field is applied in its plane \cite{magnetic}.
As it was demonstrated experimentally, fluxons can be readily trapped in
annular LJJs \cite{vernik}. In this case, the sG model takes the form

\begin{equation}
\phi _{tt}-\phi _{xx}+\sin \phi +h\sin \frac{x}{R}=-\alpha \phi _{t}-\gamma
-\Gamma \sin \omega t,  \label{sG2}
\end{equation}
where $h$ is (renormalized) magnetic field, and $R$ is the radius of the
ring. In the case of the annular junction, solutions are subject to periodic
boundary conditions, $\phi _{x}(x+2\pi R)\equiv \phi _{x}(x)$ and $\phi
(x+2\pi R)\equiv \phi (x)+2\pi N$, $N$ being the number of the trapped
fluxons (in this work, $N=1$). Comparison with the experiment shows that,
unlike the model (\ref{sG1}), the one (\ref{sG2}) is, virtually, exact \cite
{UstMal}.

We assume the spatial size of the fluxon, which is $\sim 1$ in the present
notation, to be much smaller than the circumference $L\equiv 2\pi R$ of the
annular junction. Large $L$ imposes an upper limit on the driving frequency $%
\omega $ which can support the ac-driven motion: as the fluxon's velocity
cannot exceed the maximum (Swihart) group velocity of electromagnetic waves
in LJJ, which is $1$ in our notation, Eq. (\ref{spectrum}) implies that $%
\omega \,\,_{\sim }^{<}\,\,1/L$.

A different type of ac drive for fluxons in circular LJJs was proposed in
Ref.~\onlinecite {magnetic}, viz., ac magnetic field. In this case, the
terms $h\sin \left( x/R\right) $ and $\Gamma \sin (\omega t)$ in Eq. (\ref
{sG2}) are replaced by a single one, $h\sin \left( x/R\right) \sin (\omega
t) $, which may be decomposed into two waves traveling in opposite
directions, $(1/2)[\cos (x/R-\omega t)-\cos (x/R+\omega t)]$. As it was
shown \cite{magnetic}, either traveling wave may capture a fluxon, dragging
it at the wave's phase velocity $\pm \omega R$. A similar model was proposed
in Ref. \cite{Denmark}, in which the fluxon is dragged by {\em rotating}
magnetic field. A difference of our model (which corresponds to the real
experiment reported below) is the separation between the fields that induce
the spatial modulation and ac force. The separation makes it possible to
control the dynamics in a more flexible way.

It is straightforward to derive an equation of motion for the fluxon in the
adiabatic approximation, following the lines of Refs.~\onlinecite{Bob1},%
\onlinecite{Bob2} ($\stackrel{.}{\xi }\equiv d\xi /dt$):
\begin{eqnarray}
\frac{d}{dt}\left( \frac{\stackrel{.}{\xi }}{\sqrt{1-\stackrel{.}{\xi}^{2}}}
\right) =\frac{\pi h}{4\sqrt{1-\stackrel{.}{\xi }^{2}}}\cos \frac{\xi }{R} -%
\frac{\alpha \stackrel{.}{\xi }}{\sqrt{1-\stackrel{.}{\xi}^{2}}}  \nonumber
\\
+\frac{\pi }{4}\left[ \gamma +\Gamma \sin (\omega t)\right] \,.
\label{eqnmotion}
\end{eqnarray}

For further analysis, one may assume, following Refs.~\onlinecite{magnetic}
and \onlinecite{Denmark}, that, in the lowest approximation, the fluxon is
moving at a constant velocity $\stackrel{.}{\xi _{0}}\equiv u$ belonging to
the resonant spectrum (\ref{spectrum}), so that $\xi (t)=ut+R\delta $, where
$\delta $ is a phase-locking constant. Then, the first correction to the
instantaneous fluxon's velocity, generated by the spatial modulation, can be
easily found from Eq. (\ref{eqnmotion}),

\begin{equation}
\stackrel{.}{\xi }_{1}=\left( \pi Rh/4u\right) \left( 1-u^{2}\right) \sin %
\left[ (u/R)t+\delta \right] \,.  \label{correction}
\end{equation}
The approximation applies provided that the correction (\ref{correction}) is
much smaller than the unperturbed velocity $u$, which amounts to $Rh\ll
u^{2}/\left( 1-u^{2}\right) $.

A key ingredient of the dynamical analysis is the {\it energy-balance}
equation \cite{magnetic}. In the model (3) it is based on the correction (%
\ref{correction}) to the velocity \cite{Bob1,Bob2} (while in the
above-mentioned models with ac magnetic fields \cite{magnetic,Denmark} the
approximation $\stackrel{.}{\xi }=u$ was sufficient). In the case of the
fundamental resonance, with $m=1$ in Eq. (\ref{spectrum}), i.e., $u=R\omega $%
, the energy balance yields, after a straightforward algebra,
\begin{equation}
\gamma =\frac{4\alpha R\omega }{\pi \sqrt{1-(R\omega )^{2}}}\,-\frac{\pi
h\Gamma }{8R\omega ^{2}}\left[ 1-(R\omega )^{2}\right] \cos \delta .
\label{balance}
\end{equation}

Setting $|\cos \delta |=1$ and $\gamma =0$ in Eq. (\ref{balance}) gives a
minimum ({\it threshold}) amplitude of the ac drive which can support the
fluxon's motion in the absence of the dc bias current, $\Gamma _{{\rm thr}%
}=\left( 32/\pi ^{2}h\right) \alpha R^{2}\omega ^{3}\left[ 1-(R\omega )^{2}%
\right] ^{-3/2}$. For the comparison with experimental results, the most
important consequence of Eq. (\ref{balance}) is an interval of dc bias
current density $\gamma $ in which the phase-locked ac-driven motion of the
fluxon is expected. It is produced by varying $\cos \delta $ in Eq. (\ref
{balance}) between $-1$ and $+1$:
\begin{equation}
\gamma _{1}^{-}<\gamma <\gamma _{1}^{+};\,\gamma _{1}^{\pm }\equiv \frac{%
4\alpha R\omega }{\pi \sqrt{1-(R\omega )^{2}}}\,\pm \frac{\pi R\Gamma \left[
1-(R\omega )^{2}\right] }{8(R\omega )^{2}}h.  \label{interval}
\end{equation}
Note that the size of the interval strongly depends on the driving
frequency, while in the model with the ac magnetic field \cite{magnetic}\ it
does not depend on $\omega $ at all, provided that $2\pi R\gg 1$.

Experiments have been performed with Nb/Al-AlO$_{x}$/Nb Josephson annular
junction with the mean diameter $2R=95\,\mu $m and the annulus width $5\,\mu
$m, applying the bias current $I$ and measuring the dc voltage $V$ across
the junction. The distribution of the bias current was uniform, which was
concluded from measurement of the critical current $I_{{\rm c}}$ in the
state without trapped fluxons at $H=0$. $I_{{\rm c}}$ was found to be about $%
0.9$ of its value for the small junction. The annular LJJ had the Josephson
length $\lambda _{J}\approx 30\,\mu $m and plasma frequency $50\,$GHz. Note
that these parameters imply the ratio $\sim 10$ of the junction's length $%
2\pi R$ to the fluxon's size, which is $\sim \lambda _{J}$, i.e., the
junction may indeed be regarded as a long one. The measurements were done at
the temperature $4.2\,$K, using a shielded low- noise measurement setup. The
ac driving current with the frequency $f=\omega /2\pi $ between $5$ and $26$
GHz was supplied by means of a coaxial cable ending with a small antenna
inductively coupled to the junction. The antenna was oriented coaxial to the
dc bias current supplied through the electrodes, therefore the ac magnetic
field was perpendicular to the dc magnetic field. Thus, the magnetic
component of the ac signal induced the driving force of the same type as the
bias current. The ac power levels mentioned below pertain to the input at
the top of the cryostat.

Following Ref. \onlinecite{UstMal}, trapping of a fluxon in the junction was
achieved by cooling the sample below the critical temperature ${T_{c}}
\approx 9.2\,$K for the transition of Nb into the superconductive state,
with a small dc bias current applied to the junction. At $H=0$, the fluxon
depinning current $I_{{\rm dep}}$ was found to be very small, less than $1$
of the Josephson critical current $I_{{\rm c}}$ measured without the trapped
fluxon. As a fluxon can only be trapped by junction's local inhomogeneities
in the absence of the magnetic field, this indicates at fairly high
uniformity of the junction. At low values of the field $H$, linear increase
of $I_{{\rm dep}}$ with $H$ was observed, which is well described by the
theoretical model based on Eqs.~(\ref{sG2}) and (\ref{eqnmotion}): the
zero-voltage state exists as long as the maximum fluxon's pinning force
exerted by the field-induced potential remains larger than the driving force
induced by dc bias current, which is satisfied at $|\gamma |<$ $h\,$. Note
that the fluxon depinning and re-trapping in weak external magnetic fields
were studied experimentally and analytically in Ref.~\onlinecite{UstMal}.

An evidence for the progressive ac-driven motion of the fluxon is presented
in Fig.~\ref{IVC1}a. This $I(V)$ characteristic was measured at $H=0.35\,$Oe
and $f=18.1\,$GHz. Its salient feature is two large symmetric
constant-voltage steps. The points where they intersect the zero-current
axis correspond to the fluxon moving around the junction with a nonzero
average velocity at zero dc driving force. Another remarkable feature is the
absence of any step at the zero voltage, i.e., in the present case the
fluxon cannot be trapped by the effective potential, even when the dc bias
current is small. For comparison, in Fig. \ref{IVC1}b we show the $I(V)$
curve measured at the same power and frequency of the drive, but with $H=0$.
In this case, a substantial zero-voltage step is seen, extending up to the
current $I_{0}\approx \pm 0.1$ mA. In the absence of the ac drive, the
critical current $I_{0}$ is much smaller, less than $20\,\mu $A (this
residual $I_{0}$ may be explained by small inhomogeneities of LJJ, see
above).

The conspicuous zero-voltage step in Fig. \ref{IVC1}b may be explained by
the fact that the magnetic component of the ac drive creates its own
modulated potential. This argument also helps to explain two symmetric
constant-voltage steps at $V\approx 37\,\mu $V in Fig. \ref{IVC1}b as
resonant steps supported by the ac-drive-induced modulation. Note, however,
that the latter steps do not feature zero-crossing. All the data collected
in the experiments show that the zero crossing is possible {\em solely} on
the resonant steps that occur in the presence of dc magnetic field. In other
words, the ac-driven motion of the fluxons is not possible without a
stationary spatially periodic potential. This inference is in no
contradiction with numerical results of Ref.~\onlinecite{magnetic}, where the
drive itself was spatially modulated.

Coming back to the resonant steps induced by the dc field $H$, which is the
main subject of the work, we have also measured their size vs. $H$, see Fig.~%
\ref{Istep-H}. The result is that both edge values $I_{1}^{+}$ and $%
I_{1}^{-} $ indicated in Fig.~\ref{IVC1} vary nearly linearly with $H$, up
to $H\approx 0.37\,$Oe. At still larger fields, the phase- locked ac-driven
motion of the fluxon gets interrupted in some current range (the
perturbation theory does not apply to so strong fields). These findings are
in reasonable agreement with the theoretical prediction (\ref{interval}), a
fit to which is shown by the dashed lines. This pertains to both the upward
shift of the lines $I_{1}^{\pm }(H)$ (recall that $I$ and $H$ are
proportional to $\gamma $ and $h$, respectively) and their linear change
with the magnetic field. The residual nonzero value of $I_{1}^{+}-I_{1}^{-}$
at $H=0$ matches the small non-zero-crossing step in Fig. \ref{IVC1}b. It is
noteworthy too that the current range of the zero-voltage state, $%
I_{0}^{-}<I<I_{0}^{+}$ (see Fig. \ref{IVC1}b), decreases nearly linearly
with $H$, disappearing at $H\approx 0.09\,$Oe.

Equation (\ref{interval}) also predicts a linear dependence of the step's
size on the ac-drive's amplitude $\Gamma $. Comparison with experimental
data shows an agreement in a broad power range; we do not display detailed
results of the comparison, as they display no interesting features, the
linear dependence of the range of existence of the inverse Josephson effect
vs. the amplitude of the ac signal being a common feature of all the
manifestations of this effect, including the Shapiro steps in small
junctions \cite{Barone}.

As for the dependence of the step's size on the ac-drive's frequency at a
fixed value of its power, it is hard to measure it, as variation of the
frequency inevitably entails a change in the ac power coupled to the
junction. Nevertheless, basic features reported in this work, i.e., the
zero-crossing steps at finite voltages and disappearance of the zero-voltage
state, have been observed in a broad range of the ac frequencies, starting
from about $5$ GHz and up. On the other hand, as it was mentioned above, the
condition that the moving fluxon cannot exceed the Swihart velocity $\bar{c}$
sets an upper cutoff for the frequency that can support the phase-locked
motion of fluxons. In our system, $\bar{c}$ corresponds to the dc voltage $V=%
\bar{c}\Phi _{0}/(2\pi R)\approx 80\,\mu $V, which translates, via Eq.~(\ref
{spectrum}), into the cutoff frequency $\sim 40\,$GHz for the case of the
fundamental resonance.

All the above results pertained to the fundamental resonance, $m=1$ in Eq.~(%
\ref{spectrum}). It is also easy to observe zero-crossings corresponding to
higher-orders resonances. This is illustrated in Fig.~\ref{IVC2}, showing $%
V(I)$ curves with the resonant steps generated by both the fundamental and
second-order (corresponding to $m=2$) resonances.

In conclusion, we have reported the first observation of ac-driven motion of
a topological soliton in a periodically modulated lossy medium. The
observation was made in an annular uniform Josephson junction placed into
constant magnetic field. Experimentally measured data, such as the size of
the constant-voltage step, are in good agreement with predictions of the
analytical model. The effect may take place in a broad class of nonlinear
systems and, in terms of the Josephson junctions, it may find a potential
application as a low-frequency voltage standard.

This work was partly supported by a Grant G0464-247.07/95 from the
German-Israeli Foundation. We appreciate discussions with G. Filatrella, M.
Fistul, E. Goldobin, N. Gr{\o }nbech-Jensen, and J. Niemeyer.

\begin{figure}[tbp]
\caption{Current-voltage characteristics for a fluxon trapped in the annular
Josephson junction irradiated by the ac signal with the frequency $18.1$ GHz
and power $P_{{\rm ac}}=-8$dBm. The dc magnetic field is (a) $H=0.35\,$Oe
and (b) $H=0$.}
\label{IVC1}
\end{figure}

\begin{figure}[tbp]
\caption{The critical values $I_{0}^{\pm }$ and $I_{1}^{\pm }$ of the dc
bias current, marked in Fig.~\ref{IVC1}, vs. the external dc magnetic field,
the dashed lines showing a fit to Eq.~\ref{interval}.}
\label{Istep-H}
\end{figure}

\begin{figure}[tbp]
\caption{Current-voltage characteristics for a fluxon in the annular
Josephson junction at $H=0.40\,$Oe, irradiated by the ac signal at the
frequency $10.0\,$GHz. The signal's power $P_{{\rm ac}}$ is $-3.4\,$dBm
(solid line) and $-12.4\,$dBm (dashed line). The constant-voltage steps on
the two lines correspond, respectively, to the second-order and fundamental
resonance in Eq. (\ref{spectrum}).}
\label{IVC2}
\end{figure}

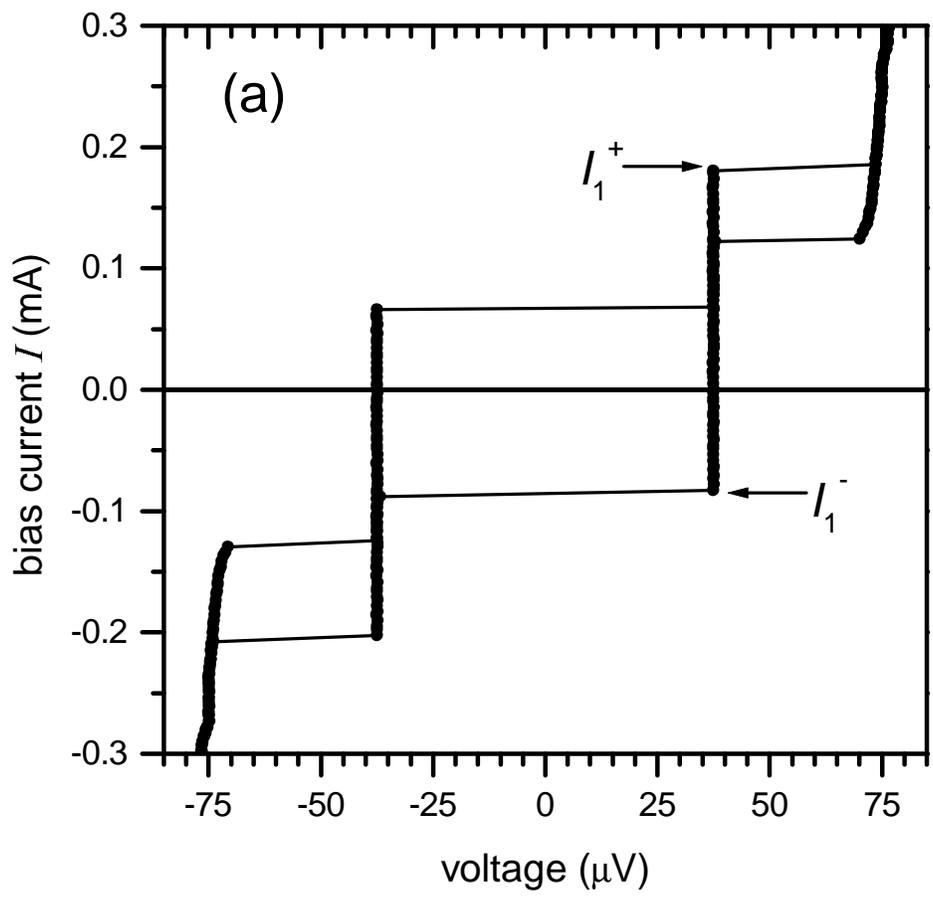
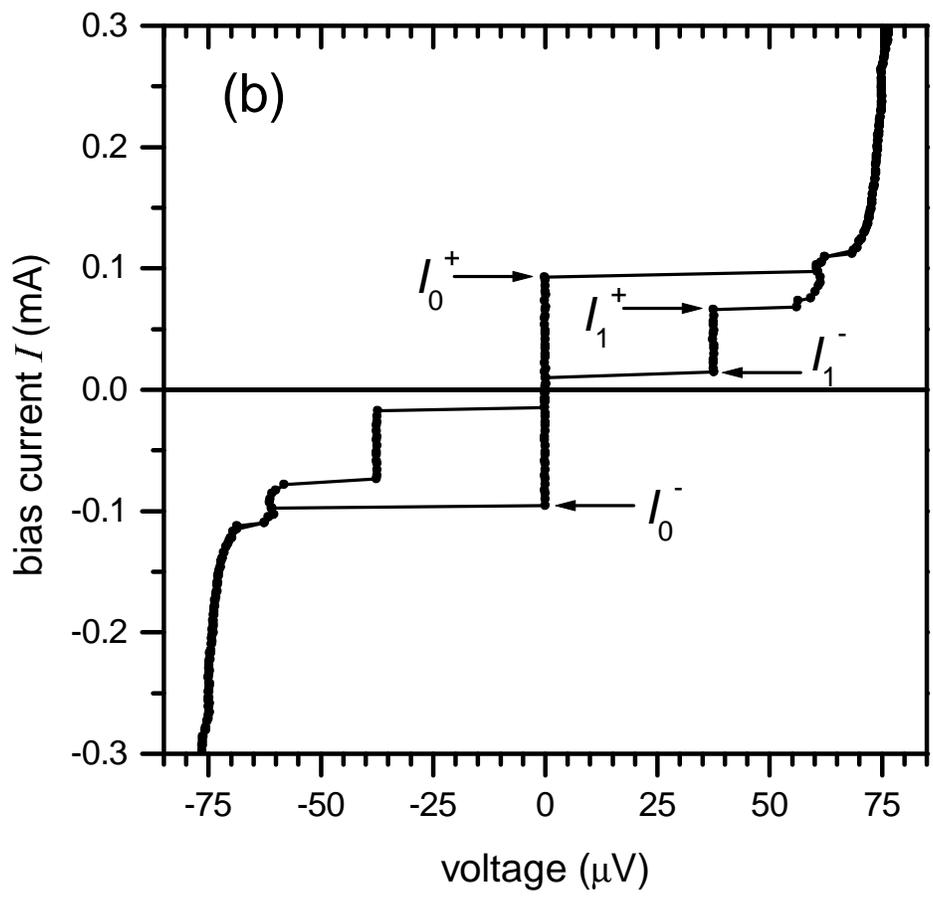
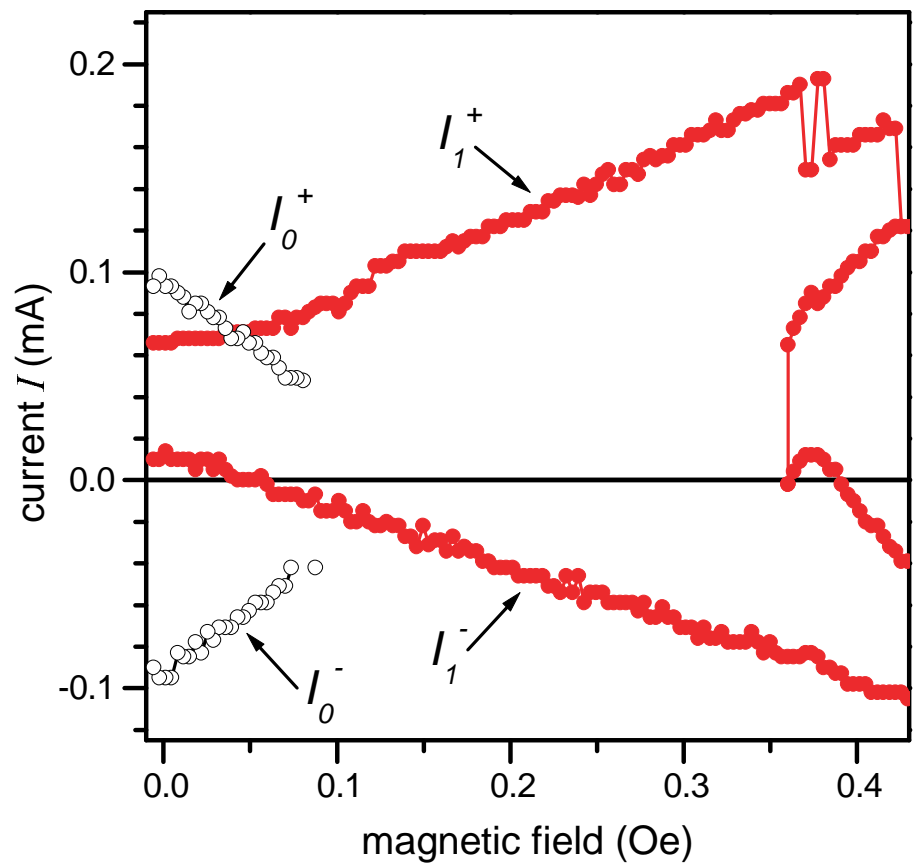
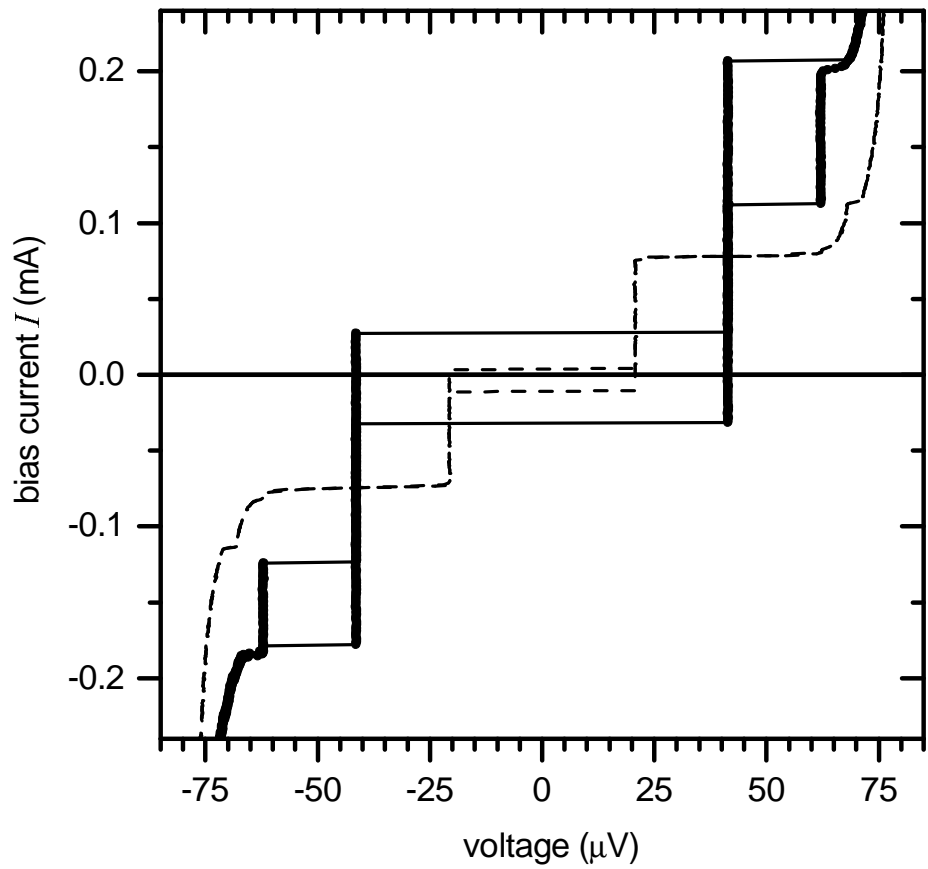
\end{document}